\documentclass[aps,prf,reprint,groupedaddress]{revtex4-2}

\usepackage{graphicx}
\usepackage{dcolumn}
\usepackage{bm}

\usepackage[utf8]{inputenc}
\usepackage[T1]{fontenc}
\usepackage{mathptmx}
\begin{document}

\title{Study on preferential concentration of inertial particles in homogeneous isotropic turbulence via Big-Data techniques}



\author{M. Obligado}
\author{A. Cartellier}
 \email{alain.catellier@univ-grenoble-alpes.fr}

\affiliation{ 
Univ. Grenoble Alpes, CNRS, Grenoble INP, LEGI, 38000 Grenoble, France
}%

\author{A. Aliseda}
\email{aaliseda@uw.edu}
\affiliation{Department of Mechanical Engineering, University of Washington, Seattle, Washington 98195-2600, USA}

\author{T. Calmant}
\author{N. de Palma}

\affiliation{ 
Univ. Grenoble Alpes, CNRS, Grenoble INP, LIG, 38000 Grenoble, France
}%

\date{\today}

\begin{abstract}

We present an experimental study of the preferential concentration of sub-Kolmogorov inertial particles in active-grid-generated homogeneous and isotropic turbulence, characterized via Vorono\"i tessellations. 

We show that the detection and quantification of clusters and voids is influenced by the intensity of the laser and high values of particles volume fraction $\phi_v$. Different biases on the statistics of Vorono\"i cells are analyzed to improve the reliability of the detection and the robustness in the characterization of clusters and voids. We do this by adapting Big-Data techniques that allow to process the particle images up to 10 times faster than standard algorithms.

Finally, as preferential concentration is known to depend on multiple parameters, we performed experiments where one parameter was varied and all others were kept constant ($\phi_v$, Reynolds number based on the Taylor length scale $Re_\lambda$, and residence time of the particles interacting with the turbulence). Our results confirm, in agreement with published work, that clustering increases with both $\phi_v$ and $Re_\lambda$. On the other hand, we find new evidence that the mean size of clusters increases with $Re_\lambda$ but decreases with $\phi_v$ and that the cluster settling velocity is strongly affected by $Re_\lambda$ up to the maximum value studied here,  $Re_\lambda = 250$.

\end{abstract}

\maketitle

\section{Introduction}

Turbulent flows laden with inertial particles are widely encountered in nature (particles dispersion in the atmosphere, rain formation, marine snow sedimentation...) and in industry (fuel or coal combustion, fluidized beds reactors, flotation techniques...). In all these configurations, inertial particles interacting with the carrier flow turbulence form high and low concentration regions leading to complex dynamics of the particles that can dominate the evolution of the flow: preferential concentration has a strong influence on the collision-coalescence of droplets in warm rain formation, among many other processes~\cite{Bateson2011}. The study of these complex flows has benefited from the fundamental theoretical background that exists for particles smaller than the Kolmogorov microscale in homogeneous isotropic turbulence (HIT). Despite simplifying assumptions, the number of independent dimensionless parameters remains large, and includes: (i) the Stokes number $St$, defined as the ratio of the particle response time $\tau_p$ to the Kolmogorov time scale of the carrier phase turbulent field, (ii) the Rouse number $Ro$, defined as the ratio of the particle terminal velocity in still fluid to the standard deviation of turbulent velocity fluctuations, (iii) the turbulent Reynolds number based on the Taylor micro-scale $Re_\lambda$, (iv) the volume fraction $\phi_v$, and the degree of polydispersity, characterized by the particle size distribution or a subset of its moments.

The role of each of these parameters on preferential concentration has not been fully characterized to date~\cite{Sumbekova}. To better understand the individual contribution of each of the flow parameters (particle inertia, gravitational crossing trajectories, carrier flow turbulence separation of scales, particle bulk collective effects, and particle-particle interactions), one parameter needs to vary while all others remain fixed. These will isolate the effect of each parameter and allow us to understand its fundamental mechanism to alter the particle dynamics in turbulence and their resulting spatial distribution and structure. 

In this manuscript, we present a series of experiments on clustering of inertial particles in a wind tunnel. All the analysis is made using Vorono\"i tessellations in 2D. Experiments were conducted so that $Re_\lambda$ and $\phi_v$ could be varied independently and, using active-grid-generated turbulence in two different modes (random and open static grid), those two parameters could be kept constant while varying the residence time of particles interacting with the turbulence. This work is unique in that it investigates the independent effect of each particle-laden turbulent flow parameter, while keeping all others constant. This decoupling of turbulence and inertial particle parameter effects is particularly important with regards to the physics of inertial particle-turbulence interactions that lead to settling velocity enhancement~\cite{aliseda2002effect} or hindering ~\cite{good}, which are still poorly understood and for which there is no community consensus on what the controlling parameters are.

In particular, the experimental results discussed here answer three open questions relative to clustering, namely: \\

\noindent \textbf{(i)} In terms of measurements reliability, are the Vorono\"i area probability density functions (PDFs) and related statistics sensitive to the average number of particles present in the images collected in experiments? This recurrent question has been considered so far by randomly removing particles from the collected sets, either from experiments or from numerical data sets. It was shown in \cite{monchaux2010preferential, monchaux2012measuring, Obligado} that the Vorono\"i analysis of clustering is weakly sensitive to sub-sampling. More specifically, relative Vorono\"i area PDFs and cluster area PDFs are nearly unchanged even if half of the particles present in the original set is randomly removed \cite{monchaux2012measuring}. Yet, in experiments and contrary to simulations, it is difficult to ascertain that all particles present in the probe volume are indeed detected, or if some bias is induced by the way images are collected and/or processed. Hence, we performed a sensitivity analysis to the average number of particles in images by collecting information in the same flow conditions and for a given optical arrangement (camera, mirrors, windows etc ...) and by varying only the laser power. \\

\noindent \textbf{(ii)} Whether or not the particles present in the probe volume are properly detected has others consequences. How the spatial distribution of particles evolves with flow conditions. depends on the reliability of the particle detection. For example, Monchaux et al. \cite{monchaux2010preferential} found that the average concentration within clusters growths with the average seeding to the power $\sim0.6$, but it was not determined if all particles were indeed detected, and therefore whether this exponent depends on the detection methods remains unclear. Similarly, the experiments in \cite{sumbekova2016clustering} presented unbiased Vorono\"i area PDFs obtained by correcting the Vorono\"i areas based on the locally averaged particle number density, to account for the vertical Gaussian intensity distribution within the laser plane. In this way, the ability to count absolute number of particles present in any given region of the flow is lost. To recover that ability, a homogeneous illumination is required to avoid using the correction routine on Vorono\"i areas. A second experimental constraint to fully understand clustering mechanism is to ensure that the number of particles detected remains proportional to the liquid volume concentration. To understand how particles are distributed in space, it is necessary to measure the exact fraction of particles present in regions with a given concentration or, at least, the fraction of particles present within clusters, in regions with average concentration and in voids, as characterized in \cite{aliseda2002effect} from line measurements with phase Doppler interpherometry (PDI) data.

\noindent \textbf{(iii)} What is the effect of carrier flow turbulence on the spatial organization of particles? Following a similar approach to \citep{sumbekova2016clustering}, where they found that the settling velocity is highly sensitive to the underlying turbulence in the carrier flow, experiments here kept constant gas velocity, constant particle volume fraction, constant particle diameter polydispersity, and constant optical arrangement, but varied the carrier flow turbulence characteristics by using an active or a passive grid. Thus, only the turbulence characteristics of the gas flow were modified to isolate the role of turbulent scale separation on the dynamics of inertial particles that are smaller than the smallest scales of the turbulence. \\

Finally, we employed novel Big-Data techniques for the analysis of massive datasets from the image acquisition stage. The image pre-processing and Vorono\"i tessellation were successfully sped up by a factor of 10 compared to  standard codes, taking advantage of the particular structure of the particle images, and the data simplification involved in the Vorono\"i that only uses the particle center location. These techniques can be easily expanded to perform more sophisticated analysis, such as cluster recognition and characterization, or other preferential concentration analysis techniques such as radial distribution function or box counting methods.       

\section{Experimental set-up}

Experiments were conducted in a large wind tunnel with a $4$~m long measurement section and a square cross-section of $0.75\times0.75$~m$^2$ that allows optical access from all sidewalls. Turbulence was generated with an active grid made of 16 rotating axes (eight horizontal and eight vertical) mounted with co-planar square blades. Each axis was driven independently with a stepping motor whose rotation rate and direction can be changed dynamically. Two different grid configurations were tested: one active and one passive. For the active mode, the motors were driven with random rotation rates and the directions were also changed randomly in time (the velocity was varied between 1 and 3 Hz, and changed randomly for time-lapses between 1 and 3s). The other protocol used is the passive open mode, where the grid was completely open, thus minimizing blockage and emulating a traditional grid. More information on the active grid and the wind tunnel can be found in \cite{Mora:aa, Obligado}. The key turbulence parameters, at the freestream velocity of $U_\infty=2.25$m/s used on most of the data presented here, are shown in table \ref{tab:table1}.

\begin{table}[h]
\begin{center}
\begin{tabular}{|c|c|c|c|c|}
\hline
 & $OG$ & $AG$ \\
\hline
$\eta$ ($\mu$m) &  $900$  & $420$    \\
\hline
$\tau_\eta$ ($ms$) & 56.0  & 12.0 \\
\hline
$Re_{\lambda}$ & 30   & 250  \\
\hline
$\lambda$ (cm) & 1.20  & 1.30 \\
\hline
$L$ (cm) & 2.4 & 6.1 \\
\hline
$u^\prime/U_\infty$ & 1.7\%  & 12.9\% \\
\hline
\end{tabular}
\caption{Turbulence parameters for a freestream velocity $U_\infty=2.25$ m/s for the active (AG) and open grid (OG) operation modes. The parameter $\eta$ is the Kolmogorov length scale and $\tau _{\eta}$ its time scale associated, $Re_{\lambda}$ the Reynolds number, $\lambda$ the Taylor micro-scale, $L$ the integral length scale (for the AG case it was calculated via the autocorrelation characteristic noise as proposed in \cite{Mora:aa}) and $u^\prime/<U_\infty>$ the fluctuations of the flow.}\label{tab:table1}
\end{center}
\end{table}

Water droplets smaller than the Kolmogorov scale of the turbulence were generated with 36 injectors (with a diameter of $0.4$~mm) arranged in a regular array (pitch $8.5$~cm along the vertical and $9$~cm along the horizontal). A $120$~bar water pump was used, with a maximum flow rate of $\sim4$ liters/minute. The volume fraction $\phi_v$ was varied in the range $\phi_v \in [0.56 \times 10^{-5} , 9.2 \times 10^{-5}]$. Drop size distributions were previously characterized \cite{sumbekova2016clustering}. The droplet distribution is polidisperse, with the most-probable diameter equal to $D_{max}=40~\mu m$ and a Sauter diameter of $D_{32}=64~\mu m$ (both values have a weak dependency with atomizing pressure and thus may change slightly in experiments where $\phi_v$ is varied). The mean Stokes number of the distribution at $U_\infty=2.25$m/s, estimated as $St=\frac{\left(D_{max}/\eta\right)^2}{36}\left(1+2 \Gamma \right)$ (where $\Gamma \sim 830$ is the density ratio between both phases) is, therefore, $St_{OG}=0.09$ and $St_{AG}=0.42$, respectively.

A vertical laser plane aligned with the freestream was formed from the top of the test section using a Powell lens with a $60^\circ$ opening angle. Powell lenses create uniform illumination conditions with a laser sheet despite the native Gaussian intensity distribution of the laser beam. The uniformity of the intensity profile is further studied on the next section. This allowed us to not correct results for particle concentration variations within the image coming from light intensity gradients (see next section). By locating the lens close to the upper window, the extent of the sheet was less than 5cm at the location where the light goes through the upper window. Therefore, the impact of droplets deposited on the top inside wall was minimized (but could not be completely eliminated). Using a double-cavity pulsed laser source (200 mJ Nd:Yag laser,  wavelength 532nm, pulse duration 4ns, Laser DualPower 200-15), we were able to vary the laser power over a significant range, namely from 30 to 170mJ per pulse (the laser output was calibrated before the experiments). The camera (Imager-Pro X) with a resolution of $1600\times1200$ pixels$^2$) equipped with an objective MIKKOR AF Micro 60 mm (F2.8d B, at full -32 - aperture) was set perpendicular to the laser sheet. To avoid any influence of deposited droplets on the tunnel wall located close to the camera, a small slit (the size of the optical aperture) was manufactured. In addition, black cardboard was placed around the slit (12cm horizontal by 10cm vertical opening) as well as at different places within the test section to minimize spurious reflections. Ambient light was totally avoided thanks to curtains surrounding the test section. Despite these precautions, some spurious reflections were still present. 

The particle imaging was done at 2.6 meters downstream the injectors. Due to optical access limitations on the top wall, the mid vertical plane of the test section was not accessible. Instead, the laser sheet was located 19 cm away from the section's wall on the camera side. This option has the advantage of diminishing the optical path length of the reflected light through the droplet field, avoiding secondary scattering by the particles. The optical axis of the camera was set at 32 cm above the test section floor (i.e. nearly at mid-height of the channel). The field of view was 20 cm along the streamwise direction and 20 cm along a vertical. The thickness of the laser for both cavities was of on the order of  $3~mm$.

The wind tunnel and the injectors were turned on until humidity and $U_\infty$ on the tunnel reached steady state. Then, water injection was stopped and the test section windows were cleaned.  Thereafter, for each run, measurements were started about $30$ seconds after the beginning of drop injection to ensure that a steady state was reached. Under conditions of strong droplet deposition on walls, this delay was reduced to a few seconds in an attempt to avoid laser propagation disturbances during the measurements. In addition, since the wind speed is modified slightly by the injection of the droplets, the gas velocity was adjusted accordingly to maintain controlled conditions (a PDI allowed to measure $U_\infty$ on the two-phase flow).

For each flow condition, 4500 frames were recorded at a 15Hz rate that corresponds to the repetition rate of  the laser cavity. The pulse duration (4ns) is small enough to freeze the motion of the droplets in the images. Since the viewing field is about 20cm long, such a low rate ensures that all images contain independent realizations of the particle-laden flow, without repeated particles in subsequent images, as long as the streamwise velocity is higher than 0.3 m/s. The present experiments were all performed at $U_\infty \geq1.1m/s$, more than fulfilling this condition. Although a 14-bit camera was used, the background noise evolved between gray levels of 80 and 120, so that the actual dynamic range was about $16384/100  \sim 160$  i.e. between 7 and 8 bits. This was taken advantage of in the Big-Data storing and processing of the images. 

\subsection{Data processing via a Big-Data approach}

In this section, we present a novel approach to data analysis via Vorono\"i tessellations that adapts Big-Data strategies to this methodology. The processing was used to detect particle centers, compute the Vorono\"i tessellations and the corresponding PDFs of cells' areas.

A collaboration with the ERODS team from the LIG (Laboratoire d'Informatique de Grenoble) led to the development of a dedicated optimized processing platform based on Big-Data, (``stream processing'' and ``cloud computing'') state-of-the-art tools.
This platform is written in Python 3 and relies on
\begin{itemize}
    \item Apache Storm, the stream processing backbone of the platform,
    \item A custom version of Yelp Pyleus, which eases the execution of Python scripts in Storm,
    \item RabbitMQ, a message queue used as storage for intermediate data,
    \item Redis, used a Rendez-Vous point in the streaming process. It is also used to share the configuration of the process between all computers.
\end{itemize}
Optionally, the platform can also be connected to InfluxDB in order to have a live feedback on the state of the whole process. The whole project has been designed to work in many configurations: scaling from 1 to hundreds of computers, possibly running heterogeneous operating systems. Indeed, the computer used for data acquisition and some analysis at the experiment run on Windows 7, whereas the cluster is Linux-based.

During the development of the platform, all computations were distributed on two Windows machines, close to the camera. This setup offered good performances, and showed that the whole process executed locally took less time than to transfer the movie files to the cluster.

\begin{figure}[t!]
\centering
\includegraphics[width=0.5\textwidth]{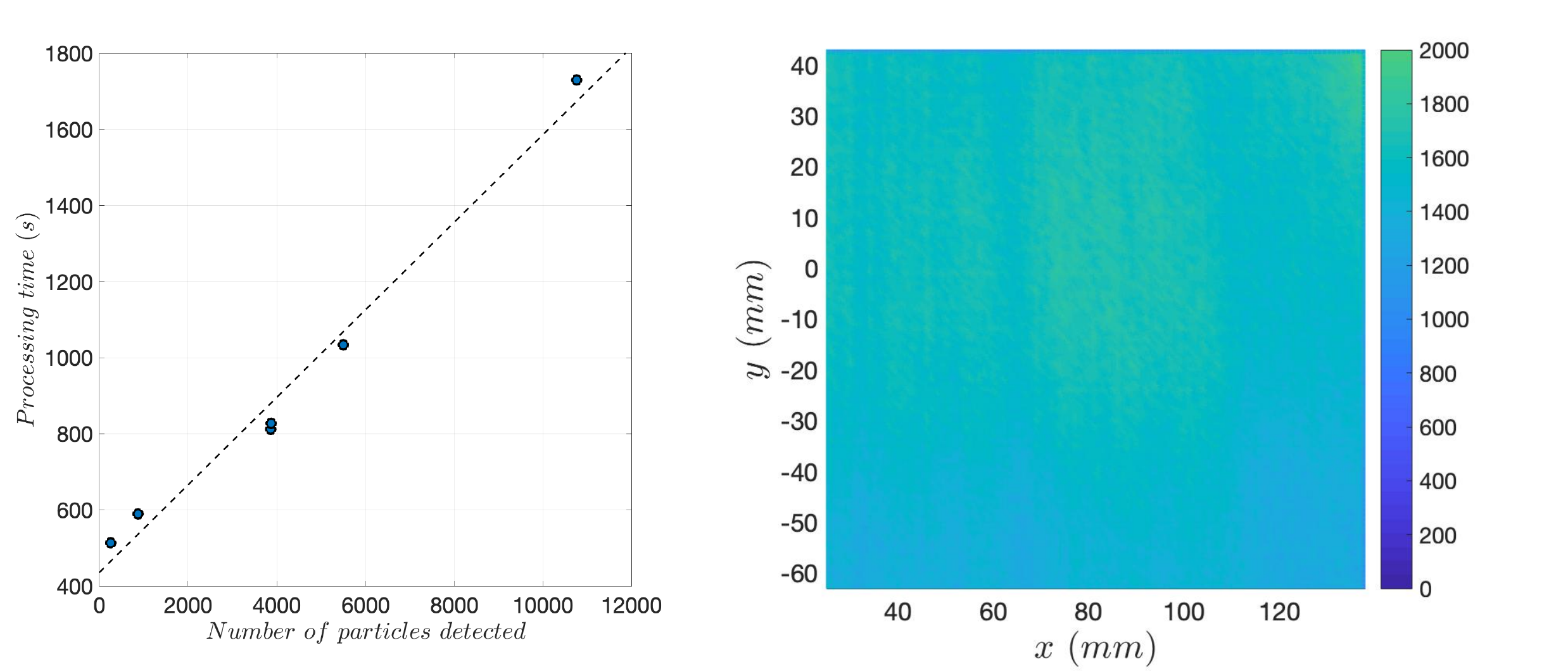}
\caption{Left: Processing time for 4500 images as a function of the mean number of particles detected per image. Right: Typical histogram of particles centers detected on the measuring window.}
\label{fig1}
\end{figure}

During this experiment, the processing was achieved on the single personal computer that retrieves the movie from the camera.
It was equipped with a 4 cores Intel Pentium processor and only of 8GB RAM.
The performance show a processing time that is linearly dependent on the total number of particles detected, as shown figure \ref{fig1}. 

In all cases, the grey-scale threshold for particle detection was set to 16 (over 255), a value that corresponds to about 6\% of the actual intensity dynamics.
Indeed, the background noise was filtered in the first stage of processing and storage of the images, so that the effective resolution was 8 bits for all images. It was found that this routine processed data up to 10 times faster than standard algorithms using the same computing power \cite{stream}.

Once particle centers were detected, the uniformity in illumination was tested for each run by mapping the mean number of particles per area (figure \ref{fig1}). It was found that the intensity is homogeneous on the whole image, except for a band where an abnormally large number of particles are detected (not shown in the image). This band was caused by a defect on the test section windows that led to reflections on the camera sensor. An ROI was defined to eliminate image boundaries as well as the vertical line from reflected light. The ROI extend was 100mm along a vertical and 110mm along the horizontal to represent one integral length-scale for the active grid and more than 4 for the open mode. Thus, only a cropped area of the image containing the uniform region was analyzed (values of $x>30~mm$ for the image in figure \ref{fig1}).

\section{Results}

\subsection{Sensitivity of Vorono\"i tessellation to the illumination intensity}

\begin{figure*}[t!]
\centering
\includegraphics[width=\textwidth]{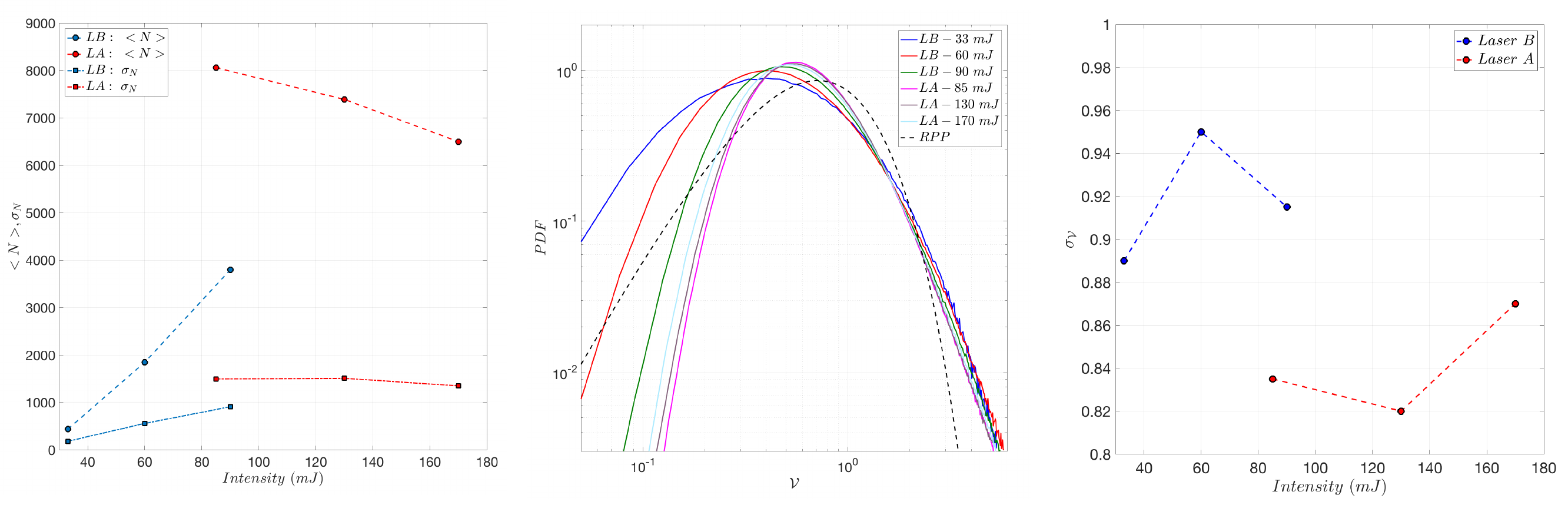}
\caption{Number of detected particles and their standard deviation vs laser intensity. Red symbols corresponds to laser A while the blue ones to laser B (\textit{left}). PDFs of normalized Vorono\"i cells areas $\cal{V}$ (\textit{center}). Standard deviation of $\cal{V}$ as a function of laser intensity (\textit{right}). All data has been obtained for a volume fraction of $\phi_v=2.47 \times 10^{-5}$ and a freestream velocity $U_\infty = 2.1$~m/s.}
\label{fig2}
\end{figure*}

Maintaining constant carrier flow conditions ($U_\infty = 2.1$~m/s, liquid flow rate 1.9~liters per minute, resulting in a particle volume fraction of $\phi= 2.47\times 10^{-5}$) and optical set-up, the laser intensity was changed at the source. The process was repeated for each of the two cavities' lasers, one spanning the range 30 to 90 mJ (referred as laser B), and the second from 85 to 170mJ (laser A). As shown in figure \ref{fig2}, for the lower illumination intensity, the mean number of particles detected per image linearly increases with the laser pulse energy. However, at the higher end of the laser illumination intensity, that number of the particles per image remains constant, showing that all particles in the flow, even the smallest ones are scattering enough light to be detected in the image. When the two lasers were operated at the same pulse energy of 90mJ, the mean number of particles detected per image was not equal, but this can be simply explained by the changes to the optical alignment required when the beam was changed from cavity A to B, with a spatial displacement of the beam. As the energy density within the laser sheet strongly depends on the alignment of the incoming beam with the Powell lens, and the light collection in the images depends on the focusing of the camera lens on the laser plane, the optical settings were clearly not identical. 

Figure \ref{fig2} center shows the PDFs of Vorono\"i cell area $\cal{V}$ (normalized with the average area). It can be observed that the standard deviation of normalized Vorono\"i cells $\sigma_{\cal{V}}$ is always above the value of a random Poisson process $\sigma_{{\cal{V}}^{RPP}}=0.53$. For 2D Vorono\"i tessellations, this parameter is an appropriate metric of the intensity of clustering~\cite{monchaux2010preferential}. In figure \ref{fig2}, it can also be appreciated that the PDFs of $\cal{V}$ present some differences in the high concentration (small area) tails of the distribution for the measurements with laser B, while the measurements with laser A present a nice collapse of those tails. The width of the distribution (quantified by the standard deviation) is  robust to intensity variations (figure \ref{fig2} right), specially at the highest illumination intensity where all the particles are captured and the number of particles in the flow (and thus the Vorono\"i cell area distribution) does not depend on the intensity of the laser or any other experimental parameter. Nevertheless, discrepancies in the Vorono\"i cell  area PDFs are always at low values of $\cal{V}$, while they all collapse for ${\cal{V}} > 1$.  Interestingly, a second (or third) crossing between the clustering and the theoretical RPP PDFs may appear at low $\cal{V}$ for the experiments in which the number of particles per image varies with experimental settings, a potentially spurious effect that should be retrospectively analyzed for many published observations. 

\begin{figure}[t!]
\centering
\includegraphics[width=0.5\textwidth]{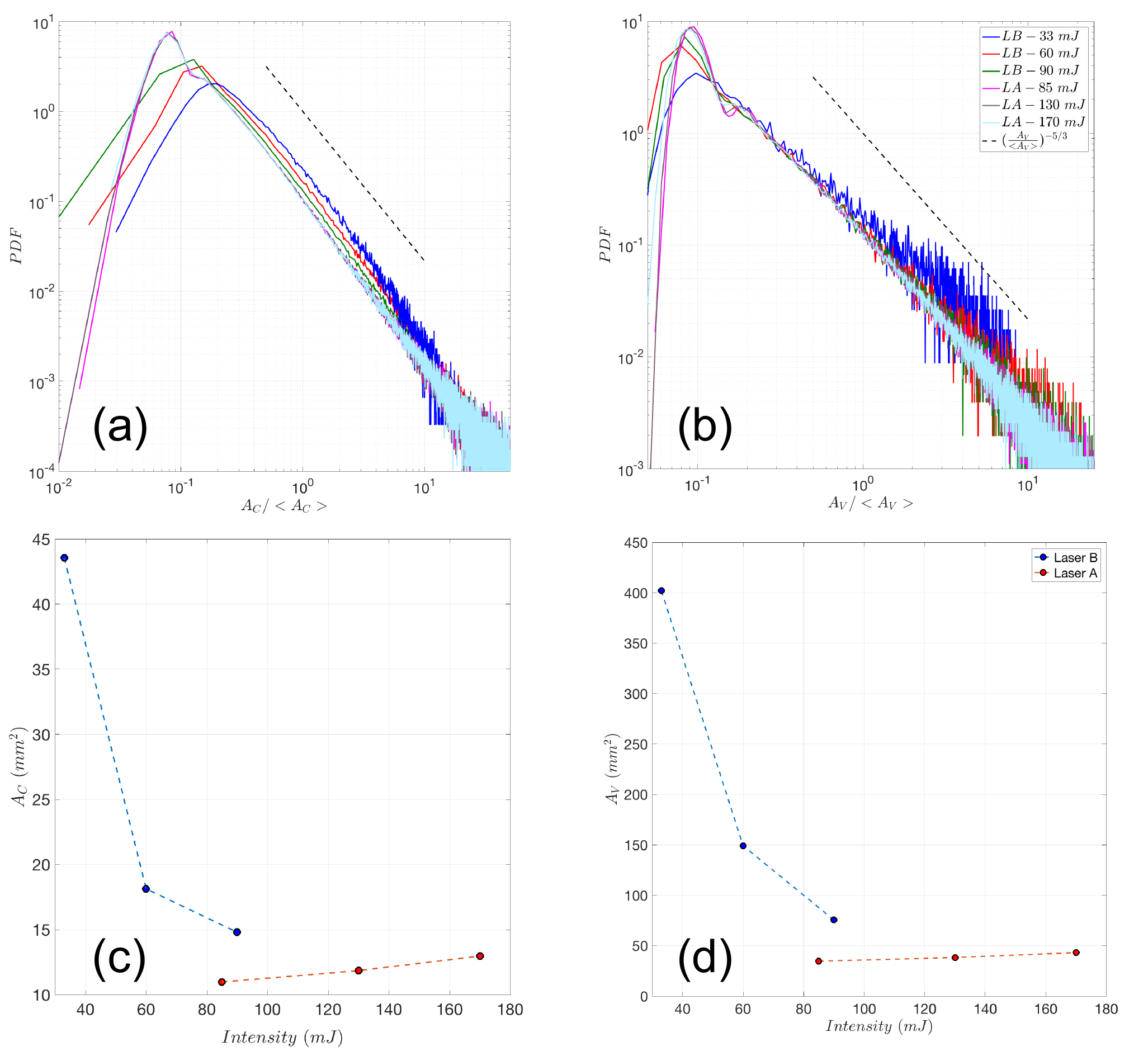}
\caption{PDF of cluster's size normalized with its mean value (a). Same figure but for voids (b). Cluster (c) and void (d) area as a function of laser intensity.}
\label{fig3}
\end{figure}

A similar trend is observed when studying the properties of clusters and voids (obtained from the PDFs of $\cal{V}$ following the method proposed by \cite{monchaux2010preferential}). Normalized size PDFs of both clusters and voids are very robust when the number of particles in the image is fully resolved (figure \ref{fig3}a\&b) while they present some inconsistencies at low clusters/voids area when the number of particles in the flow depends on experimental settings (low illumination intensity). The shape of these distributions is approximately universal, and all results presented here have a similar qualitative look. The mean values of clusters and voids areas (\ref{fig3}c\&d) also present the same trend.

Therefore, the conclusion from this study is that insufficient illumination intensity in the flow can influence the results obtained when studying clustering via Vorono\"i tessellations. Nevertheless, at high enough light intensities, all statistics remain constant, and it is possible to experimentally check the robustness of the set-up by testing different laser intensities. From now on, all reported results correspond to data collected with the set-up described for laser A at the maximum intensity of 170mJ.

\subsection{Ability to detect all particles present in the flow }

Another important question is up to which point we can detect all the particles present in the flow, for a given detection threshold. Among the expected shortcomings, the vertical attenuation of the laser sheet intensity due to droplets will eventually affect results at high values of $\phi_v$. Also, large particle concentrations can generate multiple scattering within the flow, distort the imaging and affect the detection of particles.

A second issue arises from the fact that droplets collide against the test section walls, forming a liquid layer potentially with curvature. This is particularly important on the upper wall where the liquid layer disrupts the propagation of the laser sheet. At low droplet concentration, this can manifest itself as individual droplets that remain at a constant position on the wall resulting in vertical bands of lower intensity of illumination that can be calibrated for in the analysis of the images. At higher volume fractions, the number of droplets colliding with the walls increases, and can lead to accumulation of liquid over the time of data acquisition. This situation leads to a decrease of the number of drops detected at any given pixel in the image as acquisition time proceeds, a phenomenon for which calibration can not be employed with confidence. 

\begin{figure}[t!]
\centering
\includegraphics[width=0.5\textwidth]{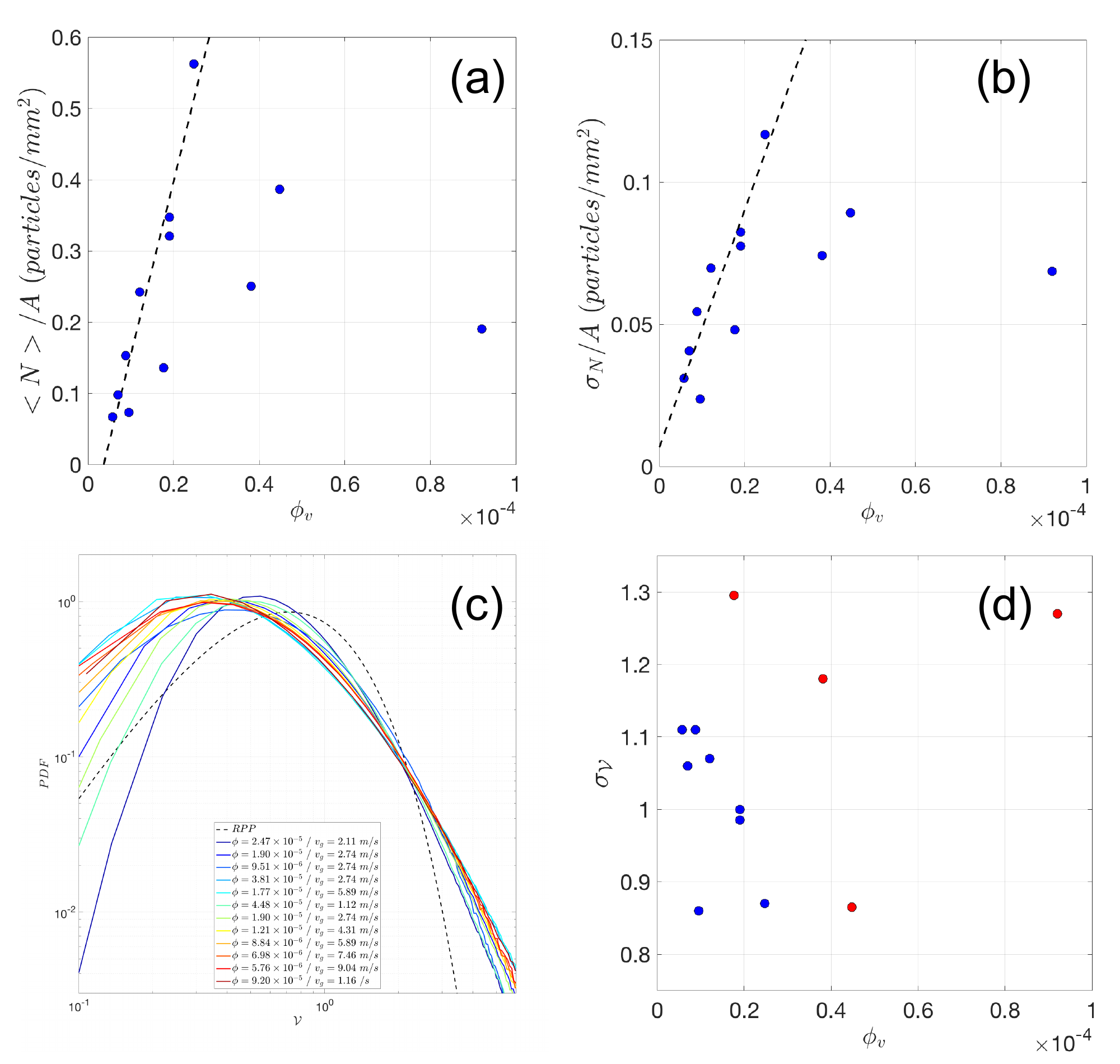}
\caption{Total number of particles detected (a) and their standard deviation (b) at constant pulse energy and different flow conditions. Corresponding PDFs of $\cal{V}$ (c) and $\sigma_{\cal{V}}$ (d). On the last figure, the events that are outliers of the straight line from figure (a) correspond to red markers.}
\label{fig4}
\end{figure}

The above two features would lead to erroneous count of particles. This is exemplified in figure \ref{fig4}a\&b obtained for a constant pulse energy (laser A at 170mJ) and for different flow conditions. The linearity between the average number of particles per image and the volume concentration holds at low values of $\phi_v$. Above values of approximately $\phi_v \sim 3\times 10^{-5}$, that linearity is lost and the average number of particles per image saturates or even decreases for very dense conditions. Note that the standard deviation of $N$ shows the same behaviour at the same threshold of $\phi_v$. On the other hand, PDFs of $\cal{V}$ and $\sigma_{\cal{V}}$ (figure \ref{fig4}c\&d) do not show any trend, nor big changes, for $\phi_v > 3 \times 10^{-5}$. The data in figures \ref{fig4}a\&b show, however, that those Vorono\"i PDFs are still not to be trusted: once more, as shown in the previous section, a new crossing at low $\cal{V}$ between the experimental and the theoretical RPP PDFs may appear, depending on the experimental conditions. This crossing, reported in various places in the literature (see for instance \cite{monchaux2010preferential,Obligado,tagawa_mercado_prakash_calzavarini_sun_lohse_2012,Sumbekova})
, may therefore be a consequence of the experimental conditions for imaging and data acquisition and not a physical phenomenon of the inertial particle concentration field.

\subsection{Clustering at different particle volume fractions and turbulence Reynolds numbers}

Preferential concentration is characterized at varying $\phi_v$ and for different turbulent Reynolds numbers, based on the Taylor length scale (as produced by the two different grid protocols indicated above, and with turbulence parameters reported in table \ref{tab:table1}).

\begin{figure}[h]
    \centering
        \includegraphics[width=\columnwidth]{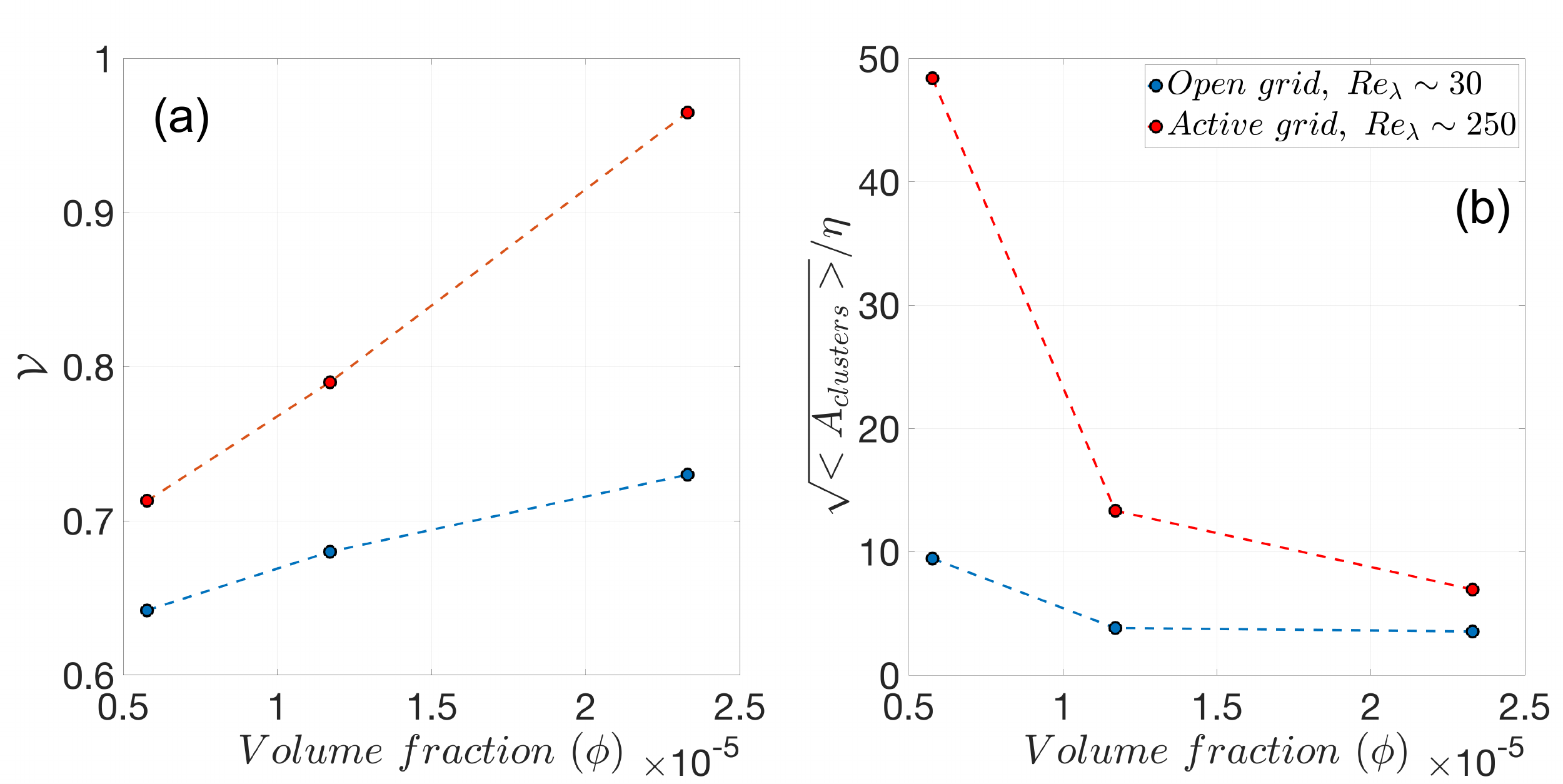}
    \caption{$\sigma_{\cal{V}}$ as a function of the volume concentration of droplets $\phi_v$ (\textit{a}). Evolution with the same parameter of the mean area of clusters ($A_{Cl}$) normalised with $\eta$ (\textit{b}).}
    \label{fig}
\end{figure}

Figure \ref{fig} shows results with varying $\phi_v$ and $Re_\lambda$, but with constant particle residence times. The high turbulence intensity case, $Re_\lambda=250$, corresponds to a Kolmogorov scale equal to $\eta=420\mu m$, while the low intensity case has a $Re_\lambda=30$ and $\eta=900\mu m$. The presence of clustering can again be observed. This data indicates the presence of clusters in the flow, increasing with both volume fraction and turbulence intensity/$Re_\lambda$, consistent with previously published results~\cite{Sumbekova}.

\begin{figure}[h]
    \centering
        \includegraphics[width=\columnwidth]{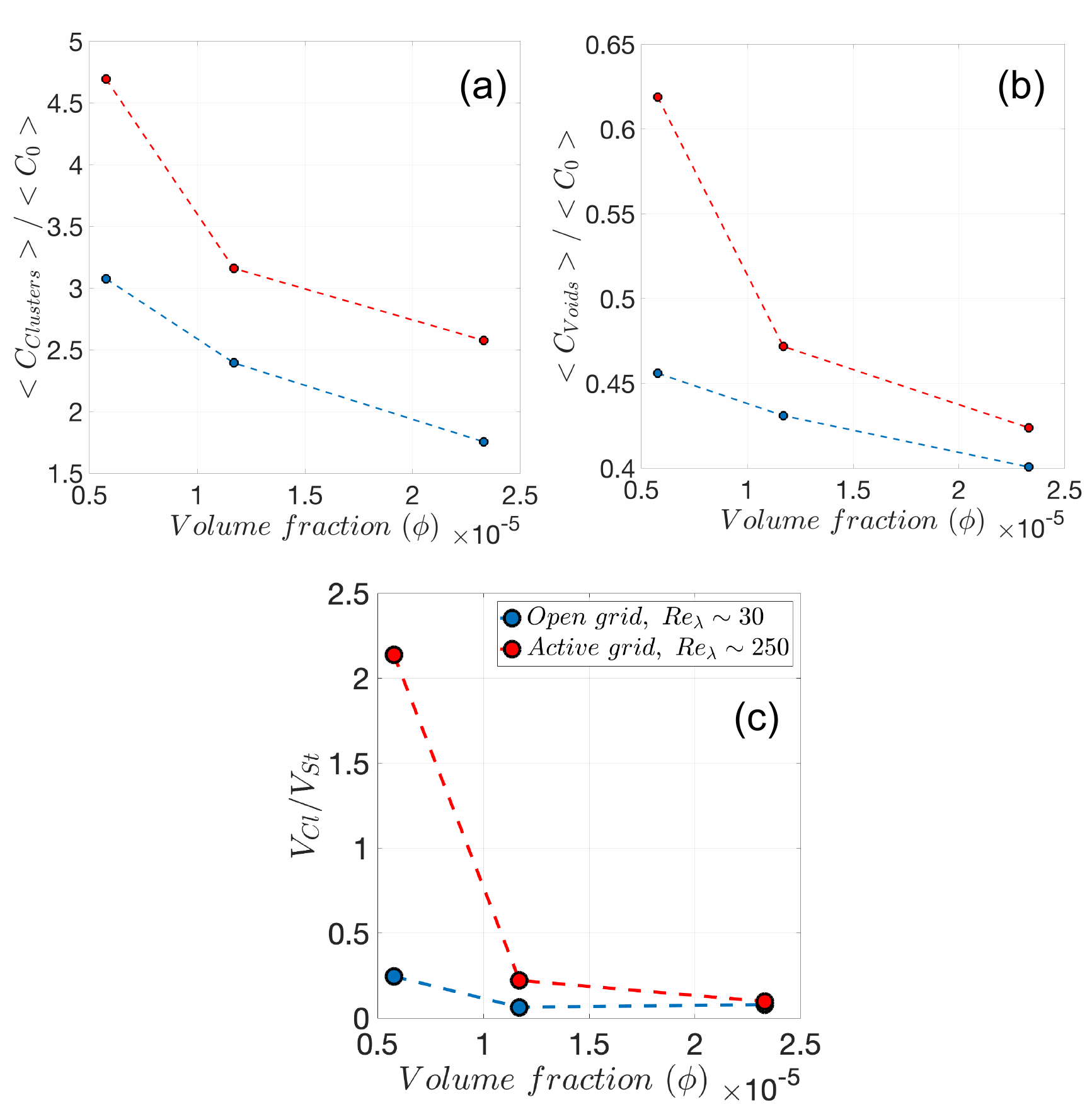}
    \caption{(\textit{a:}) Value of the mean concentration on clusters ($C_{clusters}$) normalised with the global concentration $C_0$ ($C_0=<N>/<A>$). (\textit{b:})  Same parameter but for voids.  (\textit{c:}) Cluster settling velocity, $V_{Cl}$, as a function of the same parameter. $V_{Cl}$ is defined, following the analysis from \cite{aliseda2002effect}, as $V_{Cl}=\frac{1}{60}\frac{\rho_p}{\rho_{air}}\frac{g}{\nu_{air}}<C_{Cl}><A_{Cl}>$, with $C_{Cl}$ the clusters mean volume concentration. $\rho_p$ and $\rho_{air}$ are, respectively, the particles and fluid density. The Stokes velocity is defined as $V_{St}=\sqrt{\left( \frac{4}{3} g \frac{D_{max}}{C_D} \frac{\rho_p - \rho_f}{\rho_f}\right)}$, where  the drag coefficient $C_D$ has been estimated via the Schiller-Neumann relation \cite{clift2005bubbles}. As the value of $D_{max}$ is constant, we find $V_{St}=4.56$cm for all cases. The ratios of $V_{Cl}/V_{St}$ between the random and open mode are 9, 3.5 and 1.25, respectively.}
    \label{fig7}
\end{figure}

The mean area of the clusters is found to increase with $Re_\lambda$, as also reported in~\cite{Sumbekova}, but decreases with increasing $\phi_v$, a different trend that has not been previously reported. We also found significant influence of $Re_\lambda$ on the mean concentration within clusters $<C_{Clusters}>$ and voids $<C_{Voids}>$  (figure \ref{fig7}a\&b). For instance, we find that, at the lower $Re_\lambda$, $<C_{Clusters}>/<C_0>$ is between 2 and 3 (consistent with previous measurements on the literature on similar conditions \cite{monchaux2010preferential,aliseda2002effect}). On the other hand, the larger value of Reynolds generate values of $<C_{Clusters}>/<C_0>$ that are approximately 50\% larger. Data for both low and high carrier flow Reynolds numbers follow a similar, decreasing, trend with $\phi_v$. Voids present a similar behaviour, but the difference between values of $<C_{Voids}>/<C_0>$  at different $Re_\lambda$ is significantly reduced compared to the clusters.

A similar trend is observed for the cluster settling velocity $V_{Cl}$ (see figure \ref{fig7}c where the definition of $V_{Cl}$ is detailed on the caption of this figure), suggesting that this parameter may be connected (possibly through modification of the Rouse number or the turbulence acceleration ratio, which is the ratio of the standard deviation of fluid acceleration fluctuations to gravity $\gamma ' / g$) to the changes in the settling velocity modification observed previously.

\section{Conclusions}

We have performed an experimental study on the preferential concentration of inertial particles smaller than the Kolmogorov microscale in wind-tunnel HIT. Vorono\"i tessellations, obtained using novel Big-Data techniques, allowed us to process large datasets with unprecedented efficiency, enabling better converged statistics and a wider range of parameter variations.

It is found that large values of volume fraction $\phi_v$ and deficient illumination conditions may significantly affect the PDFs of normalized Vorono\"i cells (particularly at low ${\cal{V}}$, even giving rise to an spurious crossing  between the experimental and the theoretical RPP PDFs) and the quantification of the properties of clusters is therefore severely affected. As these biases depend on the particular properties of each experimental set-up, a preliminary check should be done, changing both the intensity of the laser (to check the stability of the averaged number of particles detected $<N>$) and $\phi_v$ (to check the range where linearity between $<N>$ and $\phi_v$ is maintained), before proceeding to collect production runs for analysis. 

Nevertheless, it is found that $\sigma_{\cal{V}}$ is robust for different illumination and concentration conditions (figures \ref{fig2} \&\ref{fig4}). The reason for this is that $\sigma_{\cal{V}}$ is mostly determined by the properties and distribution of voids (as proposed in~\cite{Sumbekova}), which are much more robust to biases than the clusters.

 We also present the first experimental study of preferential concentration keeping the residence time of particles in the flow constant while varying $\phi_v$ and the carrier flow  $Re_\lambda$. Our results confirm previously reported observations, concerning the increase of $\sigma_{\cal{V}}$ and the clusters typical size when $Re_\lambda$ becomes higher \cite{Sumbekova}.

We still found, however, important differences that suggest current models of clustering need to be corrected. The analysis of cluster settling velocity found a strong dependency on $Re_\lambda$ that may be linked to the settling velocity enhancement/hindering mechanisms and switching due to the carrier turbulence characteristics controlling the values of both the Rouse number or the turbulence acceleration ratio, $\gamma ' / g$.

\section{Acknowledgments}
This work has been partially supported by the LabEx Tec21 (Investissements d'Avenir - grant agreement number ANR-11-LABX-0030). We also thanks Laure Vignal from LEGI for helping setting up the experiment.

 \bibliographystyle{plain}
\bibliography{Bib}

\end{document}